\documentclass[prd,aps,10pt,showkeys,nofootinbib,showpacs,twocolumn]{revtex4-1}

\usepackage{amsmath,amssymb,amsfonts,amsthm}

\usepackage{slashed}
\usepackage{mathtools}
\usepackage{amsfonts}
\usepackage{amssymb}
\usepackage{epsfig}
\usepackage{rotating}
\usepackage{url}
\usepackage{times}
\usepackage{color}
\usepackage{bm}
\usepackage{xcolor,colortbl}
\usepackage{hyperref}
\usepackage{enumitem}
\usepackage{anyfontsize}
\usepackage{wasysym}
\usepackage{empheq}

\newcommand{\nn}{\nonumber}

\newcommand{\be}{\begin{equation}}
\newcommand{\ee}{\end{equation}}
\newcommand{\bea}{\begin{eqnarray}}
\newcommand{\eea}{\end{eqnarray}}

\newcommand{\plk}{\mathfrak{h}}
\newcommand{\fn}{\footnote}

\newcommand{\oarX}[1]{\href{http://arxiv.org/abs/#1}{{\ttfamily #1}}}
\newcommand{\arX}[1]{\href{http://arxiv.org/abs/#1}{{\ttfamily arXiv:#1}}}

\def\barr{\begin{array}}
\def\earr{\end{array}}

\def\ben{\begin{equation}}
\def\een{\end{equation}}
\def\bs{\begin{subequations}}
\def\es{\end{subequations}}
\def\bena{\begin{eqnarray}}
\def\eena{\end{eqnarray}}

\def\im{{\rm i}}

\def\bes{\begin{eqnarray}}
\def\ees{\end{eqnarray}}

\newcommand{\dd}{\mathrm{d}}

\begin{document}

\title{Quantum analysis of the recent cosmological bounce in the comoving Hubble length}
\author{Steffen Gielen}
\email{s.c.gielen@sheffield.ac.uk}
\affiliation{School of Mathematics and Statistics, University of Sheffield,
Hicks Building, Hounsfield Road, Sheffield S3 7RH, United Kingdom}
\author{Jo\~{a}o Magueijo}
\email{j.magueijo@imperial.ac.uk}
\affiliation{Theoretical Physics Group, The Blackett Laboratory, Imperial College, Prince Consort Rd., London, SW7 2BZ, United Kingdom}

\date{\today}

\begin{abstract}
We formulate the transition from decelerated to accelerated expansion as a bounce in connection space and
study its quantum cosmology, knowing that reflections are notorious for bringing quantum effects to the fore.
We use a formalism for obtaining a time variable via the demotion of the constants 
of Nature to integration constants, and focus on a toy Universe containing only radiation and a cosmological constant $\Lambda$ 
for its simplicity. We find that,
beside the usual factor  ordering ambiguities, there is an ambiguity in the order of the quantum equation, leading to two distinct
theories: one second, the other first order. In both cases two time variables may be defined, conjugate to $\Lambda$  and 
to the radiation constant of motion. 
We make little headway with the second-order theory, but are able to produce solutions to
the first-order theory. They exhibit the well-known ``ringing'' whereby incident and reflected waves interfere,
leading to oscillations in the probability distribution even for well-peaked wave packets. We also examine in detail
the probability measure within the semiclassical approximation. Close to the bounce, the probability distribution becomes double-peaked, with one peak following a trajectory close to the classical limit but with a Hubble parameter slightly shifted  downwards, and the other with a value of $b$ stuck at its minimum. An examination of the effects still closer to the bounce, and within a more realistic model involving matter and $\Lambda$,
is left to future work. 
\end{abstract}

\maketitle

\section{Introduction}

It is not often pointed out that  the Universe has recently undergone a bounce {\it in connection space}
(not to be confused with a possible metric bounce at the Planck epoch). 
The natural connection variable in homogenous cosmological models is the inverse comoving Hubble parameter, here called $b$,
as opposed to the expansion factor $a$ in metric space (with $b=\dot a/N$ on-shell for a lapse function $N$). This is precisely 
the variable used in characterizing the horizon structure of the Universe. It is well established (see~\cite{accelexp} and references therein)
that $b$ has recently transitioned from a decreasing function of time (associated with decelerated expansion) to an increasing function of time (accelerated expansion), due to $\Lambda$ or more generally a form of dark energy taking over. If we choose the connection representation in quantum cosmology the Universe has, therefore, in the recent few billion years of its life undergone a bounce or a reflection.

Reflection is one of the best ways to highlight quantum wave-like behavior~\cite{interfreflex}, sometimes with paradoxical results~\cite{quantumreflex}. The incident and reflected waves interfere, introducing oscillations in the probability, or ``ringing'', which affects the classical limit. Such interference transforms traveling waves into stationary waves, leading to effects not dissimilar to those investigated in~\cite{randono}. Independently of this, turning points in the 
effective potential, dividing classically allowed and forbidden regions, are always regions where the WKB or semiclassical limit potentially
breaks down, revealing fully quantum behavior. The point of this paper is to initiate an investigation into this matter, specifically 
into whether the extremes of ``quantum reflection'' could ever be felt by our recent Universe. 

We base this study on recent work where a relational time (converting the Wheeler-DeWitt equation into a 
Schr\"odinger-like equation) was obtained by demoting the constants of Nature to constants-on-shell only~\cite{JoaoLetter,JoaoPaper} (i.e.,
quantities which are constant as a result of the equations of motion, rather than being fixed parameters in the action). The conjugates
of such ``constants'' supply excellent physical time variables.
This method is nothing but an extension of unimodular gravity~\cite{unimod1} as formulated in~\cite{unimod}, where the demoted
constant was the cosmological constant, $\Lambda$, and its conjugate time is Misner's volume time~\cite{misner}. 
Extensions targeting other constants (for example Newton's constant) have been considered before, notably in the context of the sequester~\cite{padilla, pad} 
in the form given in~\cite{pad1}, where the associated ``times" are called ``fluxes", or more recently in~\cite{vikman,vikman1}. 

Regarding the Wheeler--DeWitt equation in this fashion, one finds that the fixed constant solutions appear as mono-chromatic partial waves. 
By ``de-constantizing'' the constants the general
solution is a superposition of such partial waves, with amplitudes that depend on the ``de-constants''. Such superpositions can form wave packets with better normalizability properties. In this paper we investigate the simplest toy model exhibiting 
a $b$-bounce, which is a mixture of radiation and $\Lambda$, subject to the deconstantization of $\Lambda$ and a radiation variable (which can be 
the gravitational coupling $G$). The wave packets we build thus move in two alternative time variables, the description being simpler~\cite{JoaoPaper} in terms
of the clock associated with the dominant species (e.g., Misner time during Lambda domination). The $b$-bounce is the interesting epoch
where the ``time zone'' changes.

The plan of this paper is as follows. In Section~\ref{classical} we set up the classical theory
highlighting the connection rather than the metric, with a view to quantization in the connection representation (Section~\ref{quantumMSS}).
We stress the large number of decision forks in the connection representation (thus leading
to non-equivalent theories with respect to quantizations based upon the metric). Notably, beside factor-ordering issues, we have ambiguities
in the {\it order} of the quantum equation. Thus, we find two distinct theories for our toy model: one first order, and one second order. 

We seek solutions to the second-order theory in Sec.~\ref{2ndordertheory}, but encounter a number of mathematical problems that hinder 
progress. In contrast, we produce explicit solutions to the first-order theory in Section~\ref{firstordersln}, albeit at the cost of several approximations that may erase or soften important quantum behavior. Gaussian wave packets are found, and the motion of their peaks reproduces the semiclassical limit.
At the bounce they do exhibit ``ringing'' in $|\psi|^2$, as in all other quantum mechanical reflections.
However, with at least one definition of inner product and unitarity, within the semiclassical approximation this ``ringing'' disappears from the probability, as shown in Section~\ref{inner}. Nonetheless in Section~\ref{phenom} we find hints of interesting phenomenology: even within the semiclassical approximation, for a period around the bounce, the Universe is ruled by a double peaked distribution biased towards the value of
$b$ at the bounce.  This could be observable. 

Whether the features found/erased in Sections~\ref{firstordersln}--\ref{phenom} vanish or become more pronounced in a
realistic model with fewer approximations is left to future work (e.g.,~\cite{brunobounce}), as we discuss in a concluding Section.

\section{Classical theory}\label{classical}

We study a cosmological model with two candidate matter clocks, modeled as perfect fluids with equation of state parameters $w=\frac{1}{3}$ (radiation) and $w=-1$ (dark energy), respectively. In minisuperspace, these fluids can be characterized by their energy density $\rho$ or equivalently by a conserved quantity $\rho a^{3(w+1)}$. This conserved quantity is canonically conjugate to a clock variable, and hence particularly convenient to use.

Reduction of the Einstein--Hilbert action (with appropriate boundary term) to a homogeneous and isotropic minisuperspace model yields
\be
S_{{\rm GR}} = \frac{3V_c}{8\pi G}\int {\rm d}t \left(\dot{b}a^2+N a\left(b^2+k\right)\right)
\ee
where $b$ is conjugate to the squared scale factor $a^2$; varying with respect to $b$ gives $b=\dot{a}/N$, as stated above. $k=0,\,\pm 1$ is the usual spatial curvature parameter, and $V_c$ is the coordinate volume of each three-dimensional slice.

A perfect fluid action in minisuperspace can be defined by~\cite{Brown}
\be
S_{{\rm fl}} = \int {\rm d}t \left(U\dot\tau - N a^3 V_c \,\rho\left(\frac{U}{a^3 V_c}\right)\right)
\ee
where $U$ is the total particle number (whose conservation is ensured by the first term) and $\tau$ is a Lagrange multiplier. For a fluid with equation of state parameter $w$, $\rho(n)=\rho_0 n^{1+w}$ for some $\rho_0$ where $n=U/(a^3 V_c)$ is the particle number density. Now introducing a new variable $m=\frac{8\pi G \rho_0}{3V_c} (\frac{U}{V_c})^{1+w}$, conservation of $U$ is equivalent to conservation of $m$, and we can define an equivalent fluid action (see also~\cite{GielenTurok,GielenMenendez})
\be
S^{(w)}_{{\rm fl}} = \frac{3V_c}{8\pi G}\int {\rm d}t \left(\dot{m}\chi-N\frac{m}{a^{3w}}\right)\,.
\ee
The total action for gravity with two fluids is then
\bena
S_{{\rm GR}} + S^{(\frac{1}{3})}_{{\rm fl}}+S^{(-1)}_{{\rm fl}} &=& \frac{3V_c}{8\pi G}\int {\rm d}t \left[\dot{b}a^2+\dot{m}\chi_1+\dot{\Lambda}\chi_2\right.
\label{totalac}
\\&&\left.\quad-N a\left(-(b^2+k)+\frac{m}{a^2}+\frac{\Lambda}{3} a^2\right)\right]\nonumber
\eena
where we now write $m$ for the conserved quantity associated to radiation and $\Lambda/3$ for the ``cosmological integration constant'' of dark energy. (The latter is equivalent to the way in which the cosmological constant emerges in unimodular gravity~\cite{unimod}; the factor of 3 ensures consistency with the usual definition of $\Lambda$.)  We will assume that $m$ and $\Lambda$ are positive: other solutions are of less direct interest in cosmology. Classically the values of such conserved quantities can be fixed once and for all. In the quantum theory discussed below, we will only be interested in semiclassical states sharply peaked around some positive $m$ and $\Lambda$ values, even though the corresponding operators are defined with eigenvalues covering the whole real line in order to simplify the technical aspects of the theory.   

The Lagrangian is in canonical form $\mathcal{L}=p_i\dot{q}^i-\mathcal{H}$, which implies the nonvanishing Poisson brackets
\be\label{PB}
\{b,a^2\}=\{m,\chi_1\}=\{\Lambda,\chi_2\}=\frac{8\pi G}{3 V_c}
\ee
and Hamiltonian
\be
\mathcal{H} = \frac{3V_c}{8\pi G}N a\left(-(b^2+k)+\frac{m}{a^2}+\frac{\Lambda}{3} a^2\right)\,.
\label{hamiltonian}
\ee
Importantly, this Hamiltonian is {\em linear} in $m$ and $\Lambda$, and for a suitable choice of lapse given by the appropriate power of $a$, the equations of motion for $\chi_1$ and $\chi_2$ can be brought into the form $\dot\chi_i=-1$; if one allows for a negative lapse $\dot\chi_i=1$ would also be possible. Hence, in such a gauge either $\chi_1$ or $\chi_2$ are identified with (minus) the time coordinate~\cite{GielenMenendez,JoaoLetter}.

We could apply any canonical transformation to these variables, in particular point transformations from constants to functions of themselves (inducing time conjugates proportional to the original one, the proportionality factor being a function of the constants). In particular it will be convenient to introduce the canonically transformed pair
\be
\phi=\frac{3}{\Lambda}\,; \quad T_\phi= -3\frac{\chi_2}{\phi^2}
\label{canontransf} 
\ee
instead of $\Lambda$ and $\chi_2$.

Evidently, variation of Eq.~(\ref{totalac}) with respect to $N$ leads to a Hamiltonian constraint
\be
-(b^2+k)+\frac{m}{a^2}+\frac{\Lambda}{3} a^2=0
\label{hamconst}
\ee
which is equivalent to the Friedmann equation. We will think of $b$ as a ``coordinate'' and of $a^2$ as a ``momentum'' variable, and introduce the shorthand  $V(b)\equiv b^2 + k$ viewing the $b$ dependence in Eq.~(\ref{hamconst}) as a potential, whereas the $a^2$-dependent terms play the role of kinetic terms.

If we use the variables (\ref{canontransf}) from now on, we can give the two solutions to the constraint in terms of $a^2$ as
\be
a^2_\pm = \frac{\phi}{2}\left(V(b)\pm\sqrt{V(b)^2-4m/\phi}\right)
\label{sqrtsolution}
\ee
which can be seen as two constraints, linear in $a^2$, which taken together are equivalent to the original (\ref{hamconst}) which is quadratic in $a^2$. We could write this alternatively as
\be
h_\pm(b)a_\pm^2-\phi:=\frac{2a_\pm^2}{V(b)\pm \sqrt{V(b)^2-4m/\phi}} - \phi = 0
\label{linearized}
\ee
in terms of the ``linearizing'' conserved quantity $\phi$, as suggested in~\cite{JoaoLetter,JoaoPaper}. The negative sign solution in Eq.~(\ref{sqrtsolution}) corresponds to a regime in which radiation dominates ($\phi m\gg a^4$) whereas the positive sign corresponds to $\Lambda$ domination, as one can see by checking which solution survives in the $m\rightarrow 0$ or $\Lambda\rightarrow 0$ ($\phi\rightarrow\infty$) limit.

\begin{figure}[htp]
\includegraphics[scale=0.8]{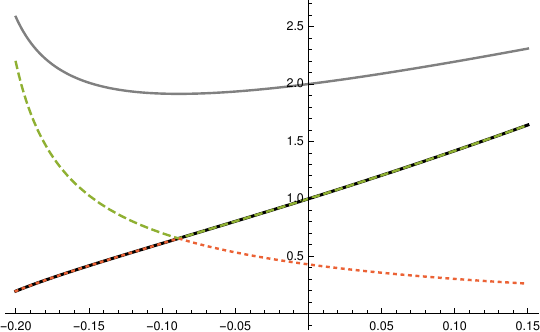}
\includegraphics[scale=0.8]{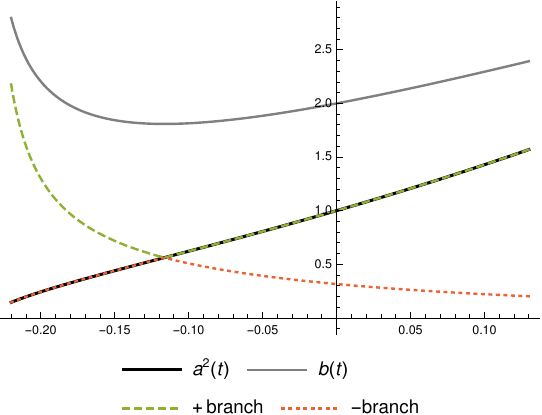}
\caption{Cosmological solutions with initial data (set at $t=0$) $a=1$, $b=2$, and $m=1.2$, (top: $k=0,\,\Lambda=8.4$, bottom: $k=1,\,\Lambda=11.4$). These follow the radiation-dominated (orange dotted) branch at small $a^2$ and the $\Lambda$-dominated (green dashed) branch at large $a^2$. The time coordinate is defined by setting $N=1/a$.}
\label{solutionplot}
\end{figure}

The equations of motion arising from Eq.~(\ref{hamconst})  can be solved numerically\footnote{Analytical solutions can be given in conformal time in terms of Jacobi elliptic functions~\cite{twosheet}.}, which shows explicitly how the classical solutions transition from a radiation-dominated to a $\Lambda$-dominated branch of Eq.~(\ref{sqrtsolution}). We plot some examples (one for $k=0$ and one for $k=1$) in Fig.~\ref{solutionplot}. Notice that the point of transition between the two branches (which is when radiation and dark energy have equal energy densities, $\phi m=a^4$) corresponds to a ``bounce'' in $b$, where $\dot{b}=0$. This bounce of course happens at a time where the Universe is overall still expanding. It happens when $V(b)=2\sqrt{m/\phi}$, or equivalently when
\be\label{b0}
b^2=b_0^2:=2\sqrt{\frac{m}{\phi}}-k\,.
\ee

It is important to realize that a linearized form of the constraints based on Eq.~(\ref{sqrtsolution}) leads to the same dynamical equations as those arising from Eq.~(\ref{hamiltonian}): for the Hamiltonian
\be
\mathcal{H}_{\pm}= \frac{3 V_c}{8\pi G}\left(a^2 - \frac{\phi}{2}\left(V(b)\pm \sqrt{V(b)^2-4m/\phi}\right)\right)
\label{linearhamilt}
\ee
we obtain
\be
\frac{{\rm d}b}{{\rm d}t}=1\,,\quad \frac{{\rm d}(a^2)}{{\rm d}t}=\phi b \left(1\pm\frac{V(b)}{\sqrt{V(b)^2-4m/\phi}}\right)\,.
\label{eom}
\ee
This form of the dynamics corresponds to a gauge in which $b$ plays the role of time and we are expressing the solution for $a^2$ in ``relational'' form $a^2(b)$. The second equation in Eq.~(\ref{eom}) can be obtained from Hamilton's equations for Eq.~(\ref{hamiltonian}) by using $\frac{{\rm d}(a^2)}{{\rm d}b}\equiv \frac{{\rm d}(a^2)}{{\rm d}t}/\frac{{\rm d}b}{{\rm d}t}$  and substituting in one of the solutions for $a^2(b)$ given by Eq.~(\ref{sqrtsolution}). Of course, this way of defining things can only ever reproduce one branch of the dynamics corresponding to one of the two possible sign choices; the equations of motion break down at the turning point $\phi m =a^4$, where one should flip from $\mathcal{H}_+$ to $\mathcal{H}_-$ or vice versa and where both the parametrization $a^2(b)$ and the gauge choice $\dot{b}=1$ in Eq.~(\ref{eom}) fail. In this sense, the ambiguities in passing from Eq.~(\ref{hamiltonian}) to the linearized form (\ref{sqrtsolution}) are related to the failure of $b$ to be a good global clock for this system, a situation frequently discussed in the literature on constrained systems~\cite{IshamRovelli}.

\section{Quantum theory}\label{quantumMSS}
Minisuperspace quantization follows from promoting the first Poisson bracket  in (\ref{PB}) to 
\begin{equation} \label{com1}
\left[ \hat{b},\hat{a}^{2}\right] =\im\frac{l_{P}^{2}}{3V_{c}},
\end{equation}%
where $l_{P}=\sqrt{8\pi G_{N}\hbar }$ is the reduced Planck length. 
Given our focus on a bounce in connection space, we choose the representation
diagonalizing $b$, so that
\be\label{a2inb}
\hat a^2=-\im\frac{l_P^2}{3V_c}\frac{\partial}{\partial b}=:-\im\plk\frac{\partial}{\partial b}
\ee
where we have introduced the shorthand $\plk$ for the ``effective Planck parameter'', as in~\cite{BarrowMagueijo}.

By choosing this representation we are making a very noninnocuous decision, leading to minimal quantum theories which 
are not dual to the most obvious ones based on the metric representation.
When implementing the Hamiltonian constraint, in the metric representation all matter contents (subject to a given theory of gravity)
share the same gravity-fixed kinetic term, with the different equations of state $w$ reflected in different powers of $a$ in the effective potential, 
$U(a)$, as is well known (e.g.,~\cite{vil-rev}).
In contrast, in the connection representation all matter fillings share the same gravity-fixed effective potential $V(b)=b^2+k$ introduced below Eq.~(\ref{hamconst}), with different 
matter components appearing as different kinetic terms, induced by their different powers of $a^2\rightarrow  -\im\plk\partial/\partial b$.

As a result, the connection representation leads to further ambiguities quantizing these theories, besides the usual 
factor-ordering ambiguities. In addition to these, we have an ambiguity in the {\it order} of 
the quantum equation (with a nontrivial interaction between the two issues). In the specific model we are studying here, we already discussed this issue for the classical theory above. We can work with the single Hamiltonian constraint (\ref{hamconst}) which is quadratic in $a^2$,
\be\label{quartic}
\frac{a^4}{\phi}-V(b)a^2+m=0\,,
\ee
with the middle term providing ordering problems; or, we can write Eq.~(\ref{quartic}) as $(a^2-a^2_+)(a^2-a^2_-)=0$ with $a_\pm^2$ given in Eq.~(\ref{sqrtsolution}), and quantize the Hamiltonian constraint written as a two-branch condition
\be\label{a2twobranch}
\hat a^2-\frac{\phi}{2}\left(V(b)\pm\sqrt {V(b)^2-4m/\phi}\right)=0\,. 
\ee
The two branches then naturally link with the monofluid prescriptions in~\cite{JoaoLetter,JoaoPaper}
when $\Lambda$ or radiation dominate (as we will see in detail later). For more complicated cosmological models in which multiple components with different powers of $a^2$ are present the situation can clearly become more complicated, with additional ambiguities in how to impose the Hamiltonian constraint. Notice also that an analogous linearization would have been possible in the metric representation, by writing Eq.~(\ref{quartic}) as $(b-b_+)(b-b_-)=0$ in terms of the two solutions for $b(a^2)$. We see no reason to expect that the resulting theories obtained by applying this procedure to either $b$ or $a^2$ would be related by Fourier transform.

We therefore have in hand two distinct quantum theories based on applying Eq.~(\ref{a2inb}) to either Eq.~(\ref{quartic}),
leading to
\be\label{quarticeq}
\left[\frac{(\hat a^2)^2}{\phi}-V(b)\hat a^2+m\right]\psi=0\,,
\ee
or to Eq.~(\ref{a2twobranch}), leading to
\be\label{a2twobrancheq1}
\left[h_\pm(b)\hat a^2-\phi\right]\psi =0
\ee
with $h_\pm(b)$ defined in Eq.~(\ref{linearized}).
One results in a second-order formulation, while the other results in a two-branch first-order formulation. 
These theories are different and there is no reason
why one (with any ordering) should be equivalent to the other. Indeed, they are not. Let us define operators
\be
D_\pm=\hat a^2-a_\pm^2(b;m,\phi)
\ee
where we work (for now) in a representation in which $m$ and $\phi$ act as multiplication operators. These operators clearly do not commute:
\be
\left[ D_+,D_-\right] \neq 0\,.
\ee
The second-order formulation, based on the constraint (\ref{quartic}), has an equation of the form
\be
:D_+ D_-:\psi=0
\ee
where the $:$ denote some conventional ``normal ordering'', for example keeping the $b$ to the left of the $a^2$.
The first-order formulation defined by Eq.~(\ref{a2twobrancheq1}) leads to a pair of equations
\be
D_+\psi=0\lor D_-\psi=0
\label{firstorderdef}
\ee
(note that an ordering prescription is implied here). 
In keeping with the philosophy of quantum mechanics, in the presence of a situation which classically corresponds to an ``OR" conjunction, 
we superpose the separate results upon quantization, so that the space of solutions is still a vector space as in standard quantum mechanics. A generic element of this solution space will satisfy neither $D_+\psi=0$ nor $D_-\psi=0$.

To understand the difference between the two types of theory, we can compare with a simple quantum mechanics Hamiltonian $H=p^2/2m + V(x)$. Quantizing the relation $E=H(p,x)$ leads to a Schr\"odinger equation that is second order in $x$ derivatives (and which, depending on the form of $V(x)$, may not be solvable analytically). Alternatively, we could replace this fixed energy relation by {\em two} conditions $p\mp{\sqrt{ 2m(E -V(x))}}=0$ linear in $p$; these would be analogous to the conditions $D_\pm = 0$ appearing in our quantum cosmology model. In the quantum mechanics case, quantizing the linear relations and taking superpositions of their respective solutions results in a set of plane-wave solutions, different to those of the second-order theory. These plane-wave solutions are interpreted as the lowest order WKB/eikonal approximation to the theory given by the initial Schr\"odinger equation. Hence, while these approaches agree in producing the same classical dynamics (away from turning points where $p$ can change sign), the two quantum theories give different predictions in terms of $\hbar$-dependent corrections to the classical limit. In quantum cosmology, we do not know which type of quantization is ``correct'' and we saw at the end of Section~\ref{classical} that the classical cosmological dynamics can be equally described by either the linear Hamiltonian (\ref{linearhamilt}) or by the original (\ref{hamiltonian}). In the quantum theory we can then follow either a first-order or a second-order approach as separate theories, with the difference between these becoming relevant at next-to-lowest order in $\hbar$. Again, we stress that this ambiguity goes beyond the issue of ordering ambiguities: it is about different {\em classical} representations of the same dynamics used as starting points for quantization. {\it The strategy proposed here is a new type of quantization procedure} compared to most of the existing quantum cosmology literature.

Indeed, no ordering prescription for the second-order formulation would lead to the total space of solutions of the first-order formulation. By choosing $:D_+ D_-:\, = D_+ D_-$, for example, the solutions of $D_-\psi=0$ would be present in the second-order formulation but not those of $D_+\psi=0$ (and vice versa). One might prefer a symmetric ordering $:D_+ D_-:\, = (D_+ D_- + D_- D_+)/2$ but the resulting equation would not be solved by solutions of either $D_-\psi=0$ or $D_+\psi=0$. If we start from a second-order formulation in which we keep all $b$ to the left,
\be\label{seconddiff}
\left(\frac{(\hat a^2)^2}{\phi}-V(b)\hat a^2+m\right)\psi=0\,,
\ee
we do not exactly recover any of the solutions of the first-order formulation, and even asymptotically (in regions in which either $m$ or $\Lambda$ dominates) we can only recover
the $D_-\psi=0$ solutions (and the radiation solutions in~\cite{JoaoLetter}). Indeed, by letting $\phi\rightarrow \infty$, Eq.~(\ref{seconddiff})
reduces to
\be
\left(-V(b)\hat a^2+m\right)\psi=0
\ee
which 
asymptotically is the same as  $D_-\psi=0$ (since $a_-^2\approx m/V(b)$ when $V(b)^2\gg 4m/\phi$). However, for $m=0$ we get
\be\label{asymptsec}
\left(\frac{\hat a^4}{\phi}- V(b) \hat a^2 \right)\psi =0
\ee
with $V(b)$ to the left of $\hat{a}^2$. Thus, we {\it cannot} factor out $\hat a^2$ on the left, to obtain 
\be
\left(\frac{1}{V(b)}\hat a^2-\phi\right)\psi=0
\ee
and so force some solutions to asymptotically match those of $D_+\psi=0$ and the pure $\Lambda$ solutions of~\cite{JoaoLetter}. The solutions of (\ref{asymptsec}) instead match those 
studied in~\cite{Ngai}. They are not the Chern--Simons state, but rather the integral of the Chern--Simons state.

From the second-order perspective, in order to reproduce the solutions of the first-order theory we would need to put the $b$ to the left or right depending on the branch we are looking at. 
The ordering in one formulation can therefore never be matched by the ordering in the other~\footnote{Apart from the forceful two-branched ordering $:D_+ D_-:\, \equiv D_+ D_-\lor D_- D_+$, of course.}.

\section{Solutions in the second-order forumlation}
\label{2ndordertheory}

In our model, as in the example of a general potential in the usual Schr\"odinger equation, the second-order theory is more difficult to solve. If we add a possible operator-ordering correction proportional to $[\hat{b}^2,\hat{a}^2]=2\im\plk\hat{b}$ to Eq.~(\ref{seconddiff}), we obtain the more general form
\be\label{seconddiff2}
\left(\frac{(\hat a^2)^2}{\phi}+\im\xi\plk b-V(b)\hat a^2+m\right)\psi=0
\ee
where $\xi$ is a free parameter (which could be fixed by self-consistency arguments; for instance, requiring the Hamiltonian constraint to be self-adjoint with respect to a standard $L^2$ inner product would imply $\xi=1$). 

We can eliminate the first derivative in Eq.~(\ref{seconddiff2}) by making the ansatz 
\be 
\psi(b,m,\phi)=e^{\frac{\im}{2\plk}\phi\left(\frac{b^3}{3}+b k\right)}\chi(b,m,\phi)
\ee
so that $\chi$ now has to satisfy
\be
\label{effectiveschr}
\left(-\frac{\plk^2}{\phi}\frac{\partial^2}{\partial b^2}+\left(m+\im \plk (\xi-1)b-\frac{\phi}{4}V(b)^2\right)\right)\chi=0
\ee
which we recognize (with $\xi=1$) as a standard Schr\"odinger equation with a (negative) quartic potential. One can write down the general solution to this problem in terms of tri-confluent Heun functions (see, e.g.,~\cite{heunbook}),
\bes
\chi&=&c_1(m,\phi)e^{-\frac{\im}{2\plk}\phi\left(\frac{b^3}{3}+b k\right)}H_T\left(\frac{m\phi}{\plk^2};-\im\frac{\phi}{\plk},-\im\frac{k\phi}{\plk},0,-\im\frac{\phi}{\plk};b\right)\nonumber
\\&&+c_2(m,\phi)e^{\frac{\im}{2\plk}\phi\left(\frac{b^3}{3}+b k\right)}H_T\left(\frac{m\phi}{\plk^2};\im\frac{\phi}{\plk},\im\frac{k\phi}{\plk},0,\im\frac{\phi}{\plk};b\right)\nonumber
\ees
where the tri-confluent Heun functions $H_T$ are normalized by defining them to be solutions to the tri-confluent Heun differential equation subject to the boundary conditions $f(0)=1$ and $f'(0)=0$. These are defined in terms of a power series around $b=0$, so that we get
\be
\psi = c_1 + c_2 + \im\,c_2\frac{k\phi}{\plk}b+\frac{m\phi(c_1+c_2)-k^2\phi^2c_2}{2\plk^2}b^2+O(b^3)\,.
\ee
These solutions could be useful for setting ``no-bounce'' boundary conditions at $b=0$ (now referring to a bounce in the scale factor), in the classically forbidden region. An immediate issue however is that tri-confluent Heun functions defined in this way diverge badly at large $b$, and are hence not very useful for studying the classically allowed region.  While they can be written down for arbitrary $\xi$, there seems to be no particular value which allows for more elementary expressions or analytical functions that are well-defined for all $b$. 

\begin{figure}[h]
\includegraphics[scale=0.8]{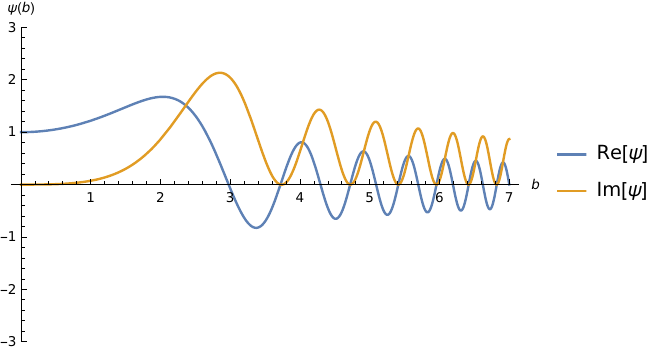}
\includegraphics[scale=0.8]{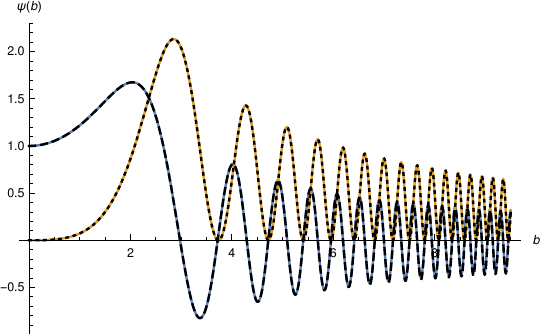}
\caption{Solutions with $m=1.2$, $k=0$, $\Lambda=8.4$ and in units $\plk=1$ (real part in blue, imaginary part in orange). The top panel shows a solution given in terms of a tri-confluent Heun function which diverges at $b\approx 6.8$. In the bottom panel we match this to the approximate solution (\ref{besselsols}) by matching the wave function and its derivative at $b=6$, which leads to a solution defined at arbitrarily large $b$. The classically allowed region is $b>b_0\approx 1.91$. This solution agrees with a numerical solution constructed from the same initial data $\psi(0)=1,\;\psi'(0)=0$ (black, dashed and dotted).}
\label{matchfig}
\end{figure}

The divergences seen in these ``analytical'' solutions are rooted in the definition of these functions as a power series around $b=0$; full numerical solutions show no such divergence but decay at large $b$. This is reassuring, but one might prefer retaining analytical expressions that can at least be valid at large $b$. In this limit, we can obtain an approximate solution by setting $m=0,\,\xi=1$, and $V(b)=b^4$ in Eq.~(\ref{effectiveschr}); the resulting differential equation has the general solution
\be
\chi=\sqrt{b}\left(c_3(m,\phi)J_{-\frac{1}{6}}\left(\frac{\phi b^3}{6h}\right)+c_4(m,\phi)J_{\frac{1}{6}}\left(\frac{\phi b^3}{6h}\right)\right)
\label{besselsols}
\ee
where $J_\nu(z)$ are Bessel functions. At large $b$, these Bessel functions have the asymptotic form
\bes
\chi &\sim& \frac{2}{b}\sqrt{\frac{3h}{\pi\phi}}\left(c_3(m,\phi)\sin\left(\frac{\phi b^3}{6h}+\frac{\pi}{3}\right)+\right.\nonumber
\\&&\left.\qquad c_4(m,\phi)\sin\left(\frac{\phi b^3}{6h}+\frac{\pi}{6}\right)\right)\,.
\ees
These asymptotic solutions are plane waves in $b^3$ modulated by a prefactor decaying as $1/b$, so they are certainly well behaved at large $b$. These large $b$ solutions can be matched to the tri-confluent Heun functions at smaller values of $b$; see Fig.~\ref{matchfig} for an example. The result of this matching agrees perfectly with a numerically constructed solution. Of course, the coefficients $c_3$ and $c_4$ in Eq.~(\ref{besselsols}) which correspond to certain initial conditions are then also only known numerically. We have no good analytical control over these solutions where they are most interesting, in the region around $b=b_0$.

If we interpret $|\psi|^2$ as a probability density, we see that this falls off as $1/b^2$ at large $b$ and so most of the probability would in fact be concentrated near the ``bounce'' $b=b_0$. One might be tempted to relate this property to the coincidence problem of cosmology, since it would suggest that an observer would be likely to find themselves not too far from equality between radiation and $\Lambda$, contrary to the naive expectation in classical cosmology that $\Lambda$ should dominate completely. Below we will compare this expectation with a more detailed calculation (and using a different measure) in the first-order theory.

We can contrast these attempts at obtaining exact solutions to the second-order theory with what would be the traditional approach in quantum cosmology, which is to resort to approximate semiclassical solutions. After all, the setup of quantum cosmology is at best a semiclassical approximation to quantum gravity.  If we start from a WKB-type ansatz $\psi(b,m,\phi)=A(b,m,\phi)e^{\im P(b,m,\phi)/\plk}$, the truncation of Eq.~(\ref{seconddiff2}) to lowest order in $\plk$ implies that
\be
\frac{1}{\phi}\left(\frac{\partial P}{\partial b}\right)^2-V(b)\frac{\partial P}{\partial b}+m=0\,,
\ee
which is the Hamilton--Jacobi equation corresponding to Eq.~(\ref{quartic}). Its solutions are $\partial P/\partial b=a^2_\pm(b;m,\phi)$ with $a^2_\pm$ as in Eq.~(\ref{sqrtsolution}), 
\be
a^2_\pm = \frac{\phi}{2}\left(V(b)\pm\sqrt{V(b)^2-4m/\phi}\right)\,,
\ee
and the general lowest-order WKB solution to the second-order theory is
\be
\psi = c_+(m,\phi)e^{\frac{\im}{\plk}\int^b {\rm d}b'\;a^2_+}+c_-(m,\phi)e^{\frac{\im}{\plk}\int^b {\rm d}b'\;a^2_-}\,.
\label{firstordersol}
\ee
On the other hand, Eq.~(\ref{firstordersol}) is already the {\em exact} general solution of the first-order theory we defined by Eq.~(\ref{firstorderdef}). These solutions are pure plane waves in the classically allowed region $|V(b)|\ge 2\sqrt{m/\phi}$ but have a growing or decaying exponential part in the classically forbidden region $|V(b)|<2\sqrt{m/\phi}$, as expected. In the next section we will discard the exponentially growing solution corresponding to $a^2_-$, but since this forbidden region is of finite extent there are no obvious normalizability arguments that mean it has to be excluded.

\section{Detailed solution in the first-order formulation}\label{firstordersln}
Needless to say, the first-order formulation is easier to solve analytically and take further. In these theories 
(e.g.,~\cite{JoaoPaper}) the general solution is a superposition of different values of ``constants''  $\bm\alpha$ of ``spatial'' 
monochromatic functions $\psi_s(b; \bm \alpha)$  (solving a Wheeler--DeWitt equation for fixed values of the $\bm\alpha$)
multiplied by the appropriate time evolution factor combining $\bm \alpha$ and their conjugates  $\bm T$.
The total integral takes the form
\be\label{gensolM}
\psi(b,\bm T)=\int  \dd{\bm \alpha}\, {\cal A}( \bm \alpha)\, \exp{\left[-\frac{\im}{\plk} \bm\alpha \cdot  \bm T  \right]}\psi_s(b; \bm \alpha)\,.
\ee
The $\psi_s$ are conventionally normalized so that in the classically allowed region
\be
|\psi_s|^2=\frac{1}{(2\pi\plk)^{D}}
\ee
where $D$ is the dimensionality of the deconstantized space, i.e., the number of conserved quantities $\bm\alpha$. The model studied in this paper corresponds to (see Eqs.~(\ref{totalac}) and (\ref{canontransf}))
\be
{\bm \alpha}=\left(\phi\equiv \frac{3}{\Lambda},m\right)\,,\;{\bm T}=\left(T_\phi,T_m =\chi_1\right)
\ee
with $D=2$.

\subsection{Monochromatic solutions}\label{monoch}

\begin{figure}
	%[h]
	\center
	\epsfig{file=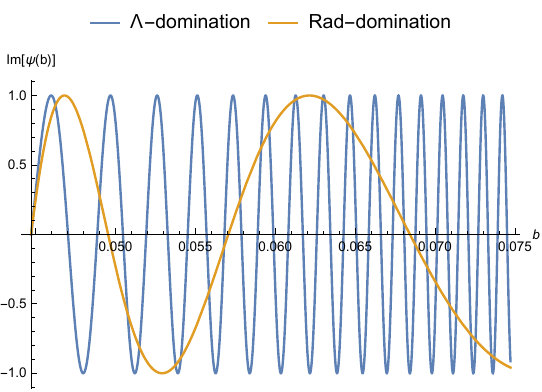,width=8.cm}
	\caption{Imaginary part of the wave functions $\psi_{s\pm}$ ($\Lambda$ branch $(+)$ in blue, radiation branch $(-)$ in orange), 
	 with $\plk =1$, $m=1$, $\phi=10^6$, defined with the 
lower limit $b_0$ (here and in the following plots $b_0\approx 0.0447$). Notice how the oscillation frequency increases/decreases with $b$ for the $\Lambda$-/radiation-dominated branches. }
	\label{Fig1}
\end{figure}

In our model, the $\psi_s(b; \bm \alpha)$ are defined to be the solutions to the two branches of Eq.~(\ref{a2twobrancheq1}), given by
\be
\psi_{s\pm}(b;\phi,m)={\cal N}  \exp{\left[\frac{\im}{\plk } \phi X_\pm  (b;\phi,m) \right]}
\label{psisdef}
\ee
with (see also Eq.~(\ref{firstordersol}))
\be
X_\pm(b;\phi ,m )=\int^b_{b_0} \dd\tilde b\,\frac{1}{2}\left(V(\tilde{b})\pm\sqrt{{V(\tilde{b})^2- 4  m/\phi }}\right),
\ee
where the integration limit is chosen to be $b=b_0$, defined in Eq.~(\ref{b0}) as the value of $b$ at the bounce.  
We plot these functions, with this choice of limits and for some particular choices of the parameters, in Fig.~\ref{Fig1}. 

\begin{figure}
	%[h]
	\center
	\epsfig{file=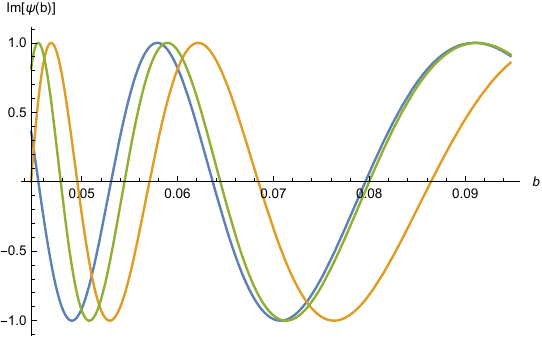,width=8.cm}
	\caption{Imaginary part of the wave functions $\psi_{s-}$ with $\plk =1$, $m=1$, and $\phi=10^6$, defined with the 
lower limit of integration $b_0$ (orange) and with lower limit infinity (green), compared with the asymptotic radiation-dominated 
wave function (blue). 
}
	\label{Fig2}
\end{figure}

We see that for $b^2\gg b_0^2$ the $+/-$ branches have
\bea
X_+(b;\phi, m)&\approx&X_\phi=\frac{b^3}{3} + kb\,,\\
X_-(b;\phi, m)&\approx&\frac{m}{\phi}X_r = \frac{m}{\phi}\int^b\frac{\dd\tilde b}{\tilde{b}^2 + k}\,,
\eea
where $X_\phi$ and $X_r$ are the corresponding functions appearing in the exponent for a model of pure $\Lambda$ (characterized by the quantity $\phi$) and a model of pure radiation. Hence, this leads to the correct limits far away from the bounce~\cite{JoaoPaper},
\bea
\psi_{s+}(b;\phi,m)&\approx &{\cal N}  \exp{\left[\frac{\im}{\plk} \phi X _\phi  (b) \right]}\,,\\
\psi_{s-}(b;\phi,m)&\approx &{\cal N}  \exp{\left[\frac{\im}{\plk} m X_r  (b) \right]}\,,
\label{xrdef}
\eea
up to a phase related to the limits of integration. This phase is irrelevant for the
$+$  wave, since $X_\phi$ diverges with $b$, so that the $b_0$ contribution quickly becomes negligible.
It does affect the $-$ wave, if we want to match with Eq.~(\ref{xrdef}) asymptotically. 
Let us assume $k=0$~\fn{The other cases are more complicated, as the Universe could become dominated by curvature before $\Lambda$ domination.}. Then, $X_-(b)\sim -\frac{1}{b}$ for large $b$, so 
in order to have agreement between Eqs.~(\ref{psisdef}) and (\ref{xrdef}) we should subtract the extra phase obtained by using $b_0$ as the lower limit
of the integral, which we denote by
\be\label{asymptphase}
\chi:= \frac{1}{\plk} \phi X_-  (\infty)\,. 
\ee
We could also take the lower limit of the integral to be $\infty$ or absorb the phase (\ref{asymptphase}) into the $-$ amplitude defined in Eq.~(\ref{superpose}),
\be
A_-\rightarrow A_-e^{\im\chi}\,.
\ee
We plot the various options for defining $\psi_{s-}$ in Fig.~\ref{Fig2}.

\begin{figure}
	%[h]
	\center
	\epsfig{file=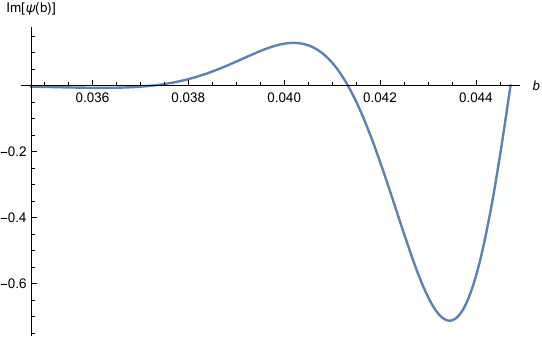,width=8.cm}
	\caption{Imaginary part of the evanescent wave function $\psi_{s-}$ valid for $b<b_0$ with $\plk =1$, $m=1$, 
	and $\phi=10^6$, defined with the lower limit $b_0$. (Strictly speaking the integrals used are only valid for $b>0$ but we
	ignore the region $b<0$.) }
	\label{Fig3}
\end{figure}

The general solution for $b>b_0$ is the superposition
\be
\psi_s(b )=A_+ \psi_{s+}(b )+  A_- \psi_{s-}(b)\,,
\label{superpose}
\ee
where we dropped the $\phi$ and $m$ labels to lighten up the notation.
In the $b<b_0$ region we have the usual evanescent wave\fn{Here we shall assume that the amplitude for tunneling into the contracting region $b<-b_0<0$ is negligible.}.
The appropriate solution (i.e., the one that is exponentially suppressed, rather than blowing up) is
\bea
\psi(b )&=&B \psi_{s-}(b )\nn\\
&=& B \exp{\left[\frac{\im }{\plk} \phi X_-  (b;\phi,m) \right]}\\
&=& B\exp{\left[\frac{ \phi }{2\plk }  \int^b_{b_0} \dd\tilde b \left(\im V(\tilde{b}) + \sqrt{{4m/\phi  - V(\tilde{b})^2  }}\right) \right]}\nn\,.
\eea
Note that the limits of integration then ensure a negative sign for the real exponential. In addition to this there is also an oscillatory factor. This solution is plotted in Fig.~\ref{Fig3}.

Our problem is now
similar to a quantum reflection problem, but with significant novelties because the medium is highly dispersive.
Usually, all we have to do is match the wave functions and their derivatives at the reflection point $b_0$ to get a fully defined wave function. 
Given that $X_+(b_0)=X_-(b_0)=0$, imposing continuity at $b=b_0$ 
requires
\be\label{cond1}
A_++A_-=B\,.
\ee
However, imposing that the first derivative of $\psi_s$ is continuous 
at $b=b_0$ produces the same
condition, given that $X'_\pm (b_0)=V(b_0)/2$. Second derivatives diverge as $b\rightarrow b_0$, as can be understood from the fact that this is a classical turning point and the monochromatic solutions are $e^{\im P(b,m,\phi)/\plk}$, where $P$ is the classical Hamilton--Jacobi function. We require as a matching condition that these divergences have the same form as we approach $b_0$ from above or below.  This leads to 
\be\label{cond2}
A_+- A_-=\im B
\ee
from a
term that diverges as $b\rightarrow b_0$. Hence
\be
\frac{A_\pm}{B}=\frac{1\pm \im}{2}\,.
\ee

For wave packets, the same conditions arise from imposing continuity
of the wave function and requiring that divergent first derivatives match, as we shall see below.

\begin{figure}
	%[h]
	\center
	\epsfig{file=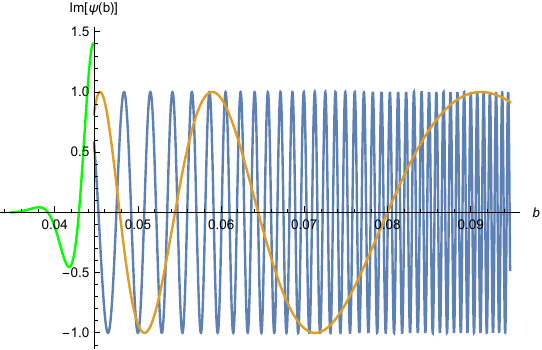,width=8.cm}
	\caption{Imaginary part of the full wave function $\psi_{s}$ normalized so as to match the asymptotic radiation-dominated expression, with 
	parameters $\plk =1$, $m=1$, 
	and $\phi=10^6$.  The incident (orange) and reflected (blue) waves, when superposed, match the evanescent wave (green) up to second derivatives in this plot.}
	\label{Fig4}
\end{figure}

Specifically, in order to match the radiation-dominated phase for the partial waves we should choose 
\bea
A_-&=&e^{-\im\chi}\,,\nn\\
B&=&\sqrt{2}e^{\im(-\chi +\pi/4)}\,,\nn\\
A_+&=&e^{\im(-\chi +\pi/2)}\,.
\label{fixedcoefs}
\eea
The resulting $\psi_s$ is plotted in Fig.~\ref{Fig4}.  Suppressing for the moment the $\bm \alpha$ label, it has the form
\bea\label{psis3terms}
\psi_s(b) &=&[A_+ \psi_{s+}(b )+  A_- \psi_{s-}(b)]\Theta(b-b_0)+\nn\\
&&\;+B\psi_{s-}(b)\Theta(b_0-b)
\eea
with the coefficients given by Eq.~(\ref{fixedcoefs}).

\subsection{Wave packets}
To construct coherent/squeezed wave packets we must now evaluate Eq.~(\ref{gensolM}) with a
factorizable state, 
\be\label{Amp}
{\cal A}(\bm\alpha)=\prod_i {\cal A}_i(\alpha_i)=
%=\sqrt{{\bf N}(\alpha_{i0},\sigma_{\alpha i})}=
\prod_i\frac{\exp{\left[-\frac{(\alpha_i-\alpha_{i0})^2}{4\sigma_{\alpha i}^2 }\right]}}
{(2\pi \sigma_{\alpha i} ^2)^{1/4}}\,.
\ee
Given Eq.~(\ref{psis3terms}), this results in
\bea\label{3pack}
\psi(b,\bm T)&=&[A_+ \psi_{+}(b,\bm T )+  A_- \psi_{-}(b,\bm T)]\Theta(b-b_0)+\nn\\
&+&B\psi_{-}(b,\bm T)\Theta(b_0-b)
\eea
with
\be\label{pmpacks}
\psi_\pm (b,\bm T)=\int  \dd{\bm \alpha}\, {\cal A}( \bm \alpha)\, \exp{\left[-\frac{\im }{\plk} \bm\alpha \cdot  \bm T  \right]}\psi_{s\pm}(b; \bm \alpha)\,.
\ee
These are the superposition of three wave packets: an incident one, coming from the radiation epoch; a reflected one, going into the $\Lambda$ epoch; and an evanescent packet in the classically forbidden region significant around the ``time'' of the bounce.

We can now follow a saddle-point approximation, as in Ref.~\cite{JoaoPaper}, which is appropriate for interpreting minisuperspace as a 
dispersive medium, where the concept of group speed of a packet is crucial. Defining the 
spatial phases $P_\pm $ from
\be
\psi_{s\pm}(b,{\bm \alpha})={\cal N} \exp{\left[\frac{\im }{\plk} P_\pm (b,{\bm \alpha})\right]}
\ee
so that
\be\label{P}
P_\pm =\phi X_\pm=\phi \int^b_{b_0} \dd\tilde b\,\frac{1}{2}\left(V(\tilde{b})\pm\sqrt{{V(\tilde{b})^2- 4  \frac{m}{\phi}}}\right)\,,
\ee
we can approximate
\be\label{expansion}
P_\pm(b,\bm \alpha)\approx P_\pm (b;\bm \alpha_0)+\sum_i\frac{\partial P_\pm }{\partial \alpha_i}\biggr\rvert_{\bm \alpha_0}(\alpha_i-\alpha_{i0})\,. 
\ee
These $P_\pm$ again correspond to the two solutions for the classical Hamilton--Jacobi function of the model, as discussed before Eq.~(\ref{firstordersol}). 
Then, for any factorizable amplitude, the wave functions (\ref{pmpacks}) simplify to
\be
\psi_\pm (b,\bm T) \approx 
e^{ \frac{i }{\plk} (P_\pm (b;\bm\alpha_0)-\bm \alpha_0\cdot \bm T)}
\prod_i \psi_{\pm i} (b,T_i)
\ee
with
\be\label{envelopes}
\psi_{\pm i}(b,T_i)=\int 
\frac{\dd\alpha_i}{\sqrt {2\pi\plk }}
\,{\cal A}_i(\alpha_i)\, e^ {-\frac{\im }{\plk} (\alpha_i-\alpha_{i0})\left(T_i- \frac{\partial P_\pm }{\partial \alpha_i}
\big\rvert_{\bm \alpha_0}
\right)}.
\ee
The first factor is the monochromatic wave centered on $\bm\alpha_0$ derived in Section~\ref{monoch}, with the 
time phases $\bm \alpha_0\cdot \bm T$ included. The other factors, $\psi_{\pm i}(b,T_i)$, describe envelopes moving with equations of motion
\be
T_i=\frac{\partial P_\pm(b,\bm\alpha )}{\partial \alpha_i}\biggr\rvert_{\bm \alpha_0}\,.
\ee
In the classically allowed region, the motion of the envelopes (and so of their peaks) reproduces the classical equations of motion
for both branches, throughout the whole trajectory, as proved in~\cite{JoaoPaper}. The packets move along 
outgoing waves whose group speed
can be set to one using the 
linearizing variable
\be
X_{\pm i}^{\rm  eff}(b)=\frac{\partial P_\pm(b,\bm\alpha )}{\partial \alpha_i}\biggr\rvert_{\bm \alpha_0}\,,
\ee
so that $T_i=X_{\pm i}^{\rm  eff}$.

Inserting (\ref{Amp}) into (\ref{envelopes}) we find that the envelopes in our case are the Gaussians
\be\label{Gauss}
\psi_{\pm i}(b,T_i)=\frac{1}{(2\pi\sigma^2_{Ti})^{1/4}} \exp\left[-\frac{ (X_{\pm i}^{\rm  eff}(b) -T_i)^2}{4 \sigma_{Ti} ^2}\right]\,,
\ee
with   $\sigma_{T_i}=\plk/(2\sigma_{\alpha i})$ saturating the Heisenberg inequality as expected for squeezed/coherent states. 

It is interesting to see that the condition (\ref{cond2}) obtained in Section~\ref{monoch} from matching divergences in the second derivative of the plane waves can be derived from the first derivative of the wave packets.  Recall that
\bea
P_\pm(b_0,\bm\alpha)&=&0\,,\nn\\
P'_\pm(b_0,\bm\alpha)&=&\phi \frac{V(b_0)}{2}=\sqrt{m\phi}\,,
\eea
to which we should add
\bea
X_{\pm i}^{\rm  eff}(b_0)&=&0\,,
\label{xefflimits}\\
\lim_{b\rightarrow b_0}\left(\sqrt{V(b)^2-4m/\phi}\,X_{\pm i}^{{\rm  eff}'  }(b)\right)&=&\mp \phi\frac{\partial(m/\phi)}{\partial\alpha_i}\,.\nn
\eea
Leaving the $A_\pm$ and $B$ undefined in Eq.~(\ref{3pack}), we then find that continuity of the wave packet at $b=b_0$ requires
\be
A_++A_-=B\,,
\ee
i.e., Eq.~(\ref{cond1}), whereas the divergent terms in the first derivative at $b_0$ agree on both sides if
\be
A_+- A_-=\im B \,,
\ee
i.e., condition (\ref{cond2}).

\begin{figure}[htp]
\includegraphics[scale=0.8]{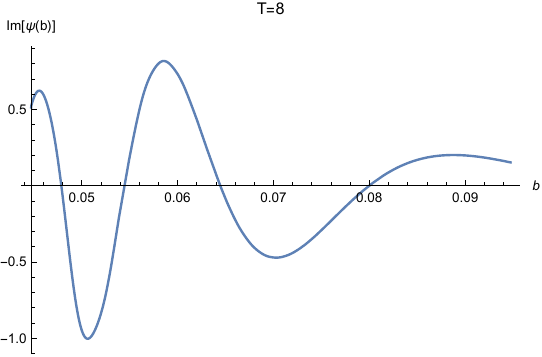}
\includegraphics[scale=0.8]{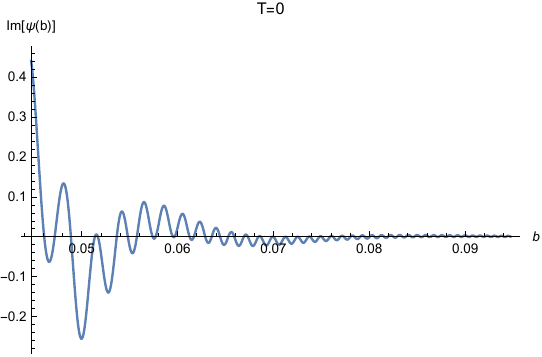}
\includegraphics[scale=0.8]{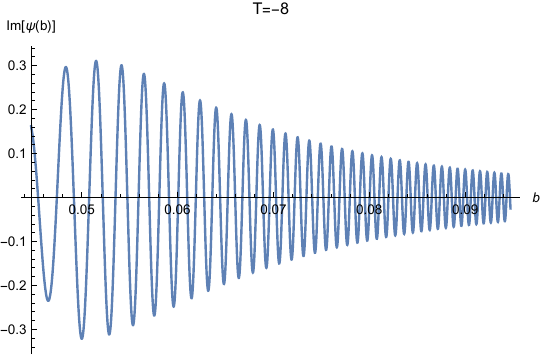}
\caption{Snapshots of the wave function in the classically allowed region $b\ge b_0$ for a wave packet with $\sigma_{Tm}=4$
at times $T_m=8,0,-8$. Note that on-shell $T_m=-(\eta-\eta_0)$, where $\eta$ and $\eta_0$ are conformal time and conformal time at the bounce, respectively. The envelope picks the right portion of the $\psi_s$, $+$ or $-$, away from the bounce. Close to the bounce, however, the $+$ and $-$ waves interfere. }
\label{reflectpsi}
\end{figure}

\begin{figure}
\includegraphics[scale=.9]{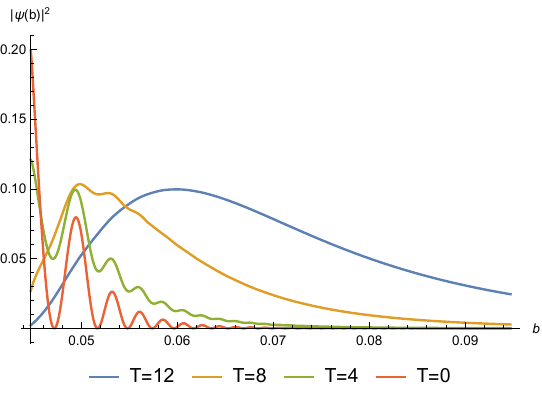}
\caption{Plot of $|\psi|^2$ for the same situation as in Fig.\ref{reflectpsi} at $T_m=12,8,4,0$.
(For the particular case of $T_m$ -- but not for a generic time -- this function is symmetric, 
so for clarity we have refrained from plotting the equivalent $T_m<0$.)}
\label{reflectprob}
\end{figure}

\subsection{Ringing of the wave function at the bounce}\label{ring}
As already studied in detail in~\cite{JoaoPaper}, the peaks of these wave packets follow the classical limit throughout 
the whole trajectory, including the bounce, assuming they remain peaked and do not interfere. They are also bona fide
WKB states asymptotically, in the sense that they have a peaked broad envelope multiplying a fast oscillating phase
(the minority clock in general will not produce a coherent packet, but we leave that matter out of the discussion here). 
The problem is that none of this applies at the bounce, where the incident and reflected waves interfere, leading to ``ringing'' in the
probability. This is an example of how the superposition of two semiclassical states is itself not a semiclassical state. 

To illustrate this point at its simplest, let us set $k=0$ and focus on the factor with the radiation time $T_m$, so that
\be\label{Xeffm}
X^{{\rm eff}}_{\pm m}=\mp\int_{b_0}^b\frac{\dd\tilde{b}}{\sqrt{\tilde{b}^4-b_0^4}}+{\rm const.}
\ee
where ${\bm \alpha}_0=(\phi_0,m_0)$, and we used that for $k=0$, Eq.~(\ref{b0}) leads to  $b_0^4=4m_0/\phi_0$.
A term constant in $b$, resulting from the dependence on $m$ in the limits of integration in (\ref{P}), can be neglected. We evaluate our wavefunctions numerically, but we note that 
in this case the integral can be expressed in terms of elliptic integrals of the first kind $F$,
\be
X^{{\rm eff}}_{\pm m}=\mp\frac{\im}{b} F(\arcsin(b/b_0);-1)+{\rm const.}
\ee
with another constant ($b$-independent) piece (which includes the constant imaginary part of the $F$ function, ensuring that the resulting $X^{{\rm eff}}_{\pm m}$ is real).

For illustration purposes, 
we then select a wave packet with $\sigma_{Tm}=4$ and follow it around the bounce
at $T_{m}=0$.  Note that on-shell $T_m=-(\eta-\eta_0)$, where $\eta$ is conformal time (shifted by $\eta_0$ so that $T_m=0$ at the bounce), 
so the conventional arrow of $T_m$ is reversed with respect to
that of $T_\phi$ or the thermodynamical arrow (see the discussion in~\cite{JoaoPaper}). In Fig.\ref{reflectpsi} we plot the wave function
away from the bounce on either side, and at the bounce.  As we see,  well away from the bounce, 
the envelope picks the right portion of the $\psi_s$ as depicted in Fig.~\ref{Fig4},  $+$ or $-$ depending on whether $T$ is positive or negative. 
Around the bounce $T=0$, however, the $+$ and $-$ waves clearly interfere (see middle plot).

As in standard reflections~\cite{interfreflex}, this interference could have implications for the probability, in the form of ``ringing''. 
We illustrate this point with the traditional $|\psi|^2$, which contains the interference cross term (but which, we stress, is not 
a serious contender for a unitary definition of probability, as we will see in the next Section). If we were to compute 
$|\psi|^2$ for the $\psi_+$ or $\psi_-$ in Fig.~\ref{Fig4} we would obtain a constant, in spite of the wave function oscillations. 
Likewise, if we dress $\psi_+$ or $\psi_-$ with an envelope, these internal beatings will not appear in the separate $|\psi|^2$. 
Close to the bounce, however, the interference between the $+$ and $-$ waves will appear as ringing in $|\psi|^2$ (see Fig.~\ref{reflectprob}) or any other measure displaying interference.  A similar construction could be made with the packets locked on to the time $T_\phi$. 

We close this Section with two words of caution. First, this ringing is probably as observable as the one associated with the mesoscopic stationary waves
described in~\cite{randono}. Indeed the two are formally related. The Chern--Simons wave function described in~\cite{randono} translates (by Fourier transform~\cite{CSHHV}) into a Hartle--Hawking stationary wave function~\cite{HH}, which is nothing but the superposition of two Vilenkin traveling waves~\cite{vil-PRD} moving in opposite directions.  The reflection studied here is precisely one such superposition in a different context and in $b$ space. The scale of the effect, however, is the same. 

Secondly, we need to make sure that the probability is indeed associated with a function (like $|\psi|^2$) containing an interference cross term, and work out the correct integration measure to obtain a unitary theory. At least with one definition of the inner product, in the semiclassical approximation the ringing
disappears, as we now show.

\section{Inner product and probability measure}\label{inner}

Usually, the inner product and probability measure are inferred from the requirement of unitarity,  i.e., the time independence of the inner
product, which in turn follows from a conserved current (see, e.g.,~\cite{vil-rev,vil-PRD}). As explained in Ref.~\cite{JoaoPaper}, in monofluid situations
this leaves us with three equivalent  definitions, which we first review.

\subsection{Monofluids}\label{innermono}
For a single fluid with equation of state parameter $w$, the first-order version of the Hamiltonian constraint leads to a dynamical equation that can be written as
\be\label{outeq}
\left((b^2+k)^{\frac{2}{1+3w}}\frac{\partial}{\partial b}+
 \frac{\partial}{\partial T}
\right)\psi =: \left(\frac{\partial}{\partial X}+
 \frac{\partial}{\partial T}
\right)\psi =0
\ee
with $T$ dependent on $w$ and 
\begin{equation}
X=\int\frac{{\rm d}b}{(b^2+k)^{\frac{2}{1+3w}}}.
\end{equation}
From such an equation we can infer a current $j^X=j^T=|\psi|^2$ satisfying the  conservation law
\be
\partial_X j^X +\partial_T j^T=0\,.
\ee
The inner product can then be defined as
\be\label{innX}
\langle\psi_1|\psi_2  \rangle=\int {\rm d}X \psi_1^\star(b(X),T)\psi_2(b(X),T)
\ee
with unitarity enforced by current conservation:
\be
\frac{\partial}{\partial T}\langle\psi_1|\psi_2  \rangle=-\int {\rm d}X \frac{\partial}{\partial X}(\psi_1^\star(b,T)\psi_2(b,T))=0\,.
\ee
For this argument to be valid without the introduction of boundary conditions as in, e.g.,~\cite{GielenMenendez}, here and in the following we must assume that $X(b)$ takes values over the whole real line and is monotonic. This is true for many cases including the ones studied here, namely, radiation and $\Lambda$ with $k=0$ (and also in the case of dust with $k=0$, studied in~\cite{brunobounce}). 
We have then established that a useful integration measure for monofluids is
\be
\dd\mu(b)=\dd X= \frac{\dd b}{\left|(b^2+k)^{\frac{2}{1+3w}}\right|}.
\ee
The normalizability condition $|\langle\psi |\psi   \rangle|=1$ supports using this measure to identify the probability. 
Given the particular form of the general solution for monofluids,
\be\label{wavepacketmono}
\psi (b,T)=\int \frac{\dd\alpha}{\sqrt {2\pi\plk }} {\cal A}(\alpha) \exp{\left[\frac{\im }{\plk}\alpha (X(b) - T)  \right]}\,,
\ee
we can write (\ref{innX}) in the equivalent forms
\bea
\langle\psi_1|\psi_2  \rangle&=&\int \dd T \;\psi_1^\star(b,T)\psi_2(b,T)\,,\label{innT}\\
\langle\psi_1|\psi_2  \rangle &=&\int \dd\alpha \; {\cal A}_1^\star (\alpha) {\cal A}_2(\alpha)\, .\label{innalpha}
\eea

\subsection{Multifluids with no bounce}\label{innermulti}

Unfortunately, not all of this construction generalizes to the transition regions of multifluids, 
where an ``$X$'' variable can be defined, but in general depends on $\alpha$ as well as $b$ (even putting aside that 
there may be multibranch expressions if there is a bounce, a matter which we ignore at first). 

We {\it may} propose that the inner product in a general multifluid setting be defined by the generalization of 
(\ref{innalpha}),
\be\label{innalpha1}
\langle\psi_1|\psi_2  \rangle=\int \dd{\bm \alpha} \; {\cal A}_1^\star ({\bm \alpha}) {\cal A}_2({\bm \alpha})
\ee
which,
by construction, is time-independent, and so unitarity is preserved. However, since $\psi_s$ in (\ref{gensolM})
is not a plane wave in some $X(b)$, its
expressions in terms of integrals in $b$ and $\bm T$ will not generally take the forms (\ref{innX}) and (\ref{innT}). 
For example,
\bea
\langle\psi_1|\psi_2  \rangle&=&\int \dd{\bm T}\, \dd{\bm T' }\, \psi_1^\star(b,\bm T)\psi_2(b,\bm T')K(b,\bm T-\bm T')\nn
\eea
with 
\bea
K(b,\bm T-\bm T')
&=&\int \dd\bm \alpha \frac{e^{-\frac{\im}{\plk}\bm\alpha\cdot(\bm T-\bm T')}}{(2\pi\plk)^{2D}|\psi_s(b,\bm\alpha)|^2}\nn
\eea
so we recover Eq.~(\ref{innT}) if and only if $\psi_s$ is a pure phase\fn{As we saw in Eq.~(\ref{3pack}), in the case of a bounce $\psi_s$ must be chosen to be a superposition of the solutions $\psi_{s+}$ and $\psi_{s-}$ in the classically allowed region, so this condition is not met.}.
Even if $\psi_s$ is a pure phase, we would not be able to recover a form like Eq.~(\ref{innX}) which would
require $\psi_s$ to be a plane wave in some $X$ only dependent on $b$. In general, the kernel $K(b-b',\bm T)$ for the $X$ inner product 
will not be diagonal, inducing an interesting new quantum effect\fn{This would in principle interact with ``ringing'' in a case where incident
and reflected waves interfere.}.

\subsection{Semiclassical measure}\label{innersemi}

With the proviso that {\it this might erase important quantum information}, 
the discussion simplifies within 
the wave packet approximation (already used in Sec.~\ref{ring}).
Then, the calculation of the measure in terms of $b$ is straightforward. We call the measure thus inferred 
 the semiclassical measure, since it fully erases quantum effects, as we shall see. 

Still ignoring the bounce (and so the double-branch) setup, 
we can regard minisuperspace for multifluids as a dispersive medium with the single
dispersion relation~\cite{JoaoPaper}
\be
\bm \alpha \cdot \bm T-P(b,\bm \alpha)=0\,.
\ee
If  the 
amplitude ${\cal A}({\bm \alpha})$ is factorizable and sufficiently peaked around ${\bm \alpha}_0$ we
can Taylor expand $P$ around $\bm\alpha_0$ to find
\be\label{approxwavepack}
\psi\approx 
e^{ \frac{\im}{\plk}(P(b;\bm\alpha_0)-\bm \alpha_0\cdot \bm T)}
\prod_i \psi_i(b,T_i)
\ee
with (cf.~Eq.~(\ref{envelopes}))
\bea
\psi_i(b,T_i)&=&\int \dd\alpha_i\,{\cal A}(\alpha_i) \frac{e^ {-\frac{\im}{\plk} (\alpha_i-\alpha_{i0})(T_i-X^{\rm eff}_i
)}}{\sqrt{2\pi\plk}}\,,\\
X^{\rm eff}_i&=&\frac{\partial P}{\partial \alpha_i}{\Big|}_{\alpha_{i0}}\,.
\eea
Then,  for 
the space of all of the functions with an ${\cal A}(\bm\alpha)$ factorized as $ {\cal A}(\bm\alpha)=\prod_{i=1}^D {\cal A}_i(\alpha_i)$
and peaked around the {\it same} ${\bm\alpha}_0$, 
the definition (\ref{innalpha1}) simplifies to
\be
\langle\psi_1|\psi_2  \rangle= \prod_{i=1}^{D}\int \dd{\alpha_i} \; {\cal A}_{i1}^\star (\alpha_i) {\cal A}_{i2}(\alpha_i) 
\ee 
and is equivalent to\fn{The amplitude functions in this space, we stress, are not necessarily Gaussian and, if Gaussian, do not necessarily have to have the same variance, but they must all peak around the same $\bm\alpha_0$ for the argument to follow through.}
\be
\langle\psi_1|\psi_2  \rangle=\prod_{i=1}^{D} \int \dd{X^{{\rm eff}}_i}  \psi _{i1}^\star (b,T_i) \psi_{ i2}(b,T_i)
\label{differentinnerprod}
\ee 
with $\dd X^{\rm eff}_i=(\dd X^{\rm eff}_i/\dd b)\dd b$. 
Hence, in this approximation, in the presence of multiple times the probability factorizes,
\be
{\cal P}(b,\bm T)=\prod_{i=1}^{D} {\cal P}_i(b,T_i)\,,
\ee
and each factor is normalized with respect to the measure
\be
\dd\mu_i(b)=\dd X^{{\rm eff}}_i
\ee
which we identify as the semiclassical probability measure.  This normalization implies that each ${\cal P}_i(b,T_i)$ can itself be seen as a probability distribution for $b$ at a particular value of $T_i$, with unspecified values for the other times.

\subsection{Case of a bounce}\label{innerbounce}

In our case $D=2$, so the wave function is the product of two independent factors, one for $m$ and one for $\phi$ (and their respective clocks). 
The fact that there is a bounce in $b$ adds an
extra complication. Indeed, each factor is the superposition of three terms:
the incident ($-$) wave, the reflected ($+$) wave, and the evanescent wave. A crucial novelty is that 
$X^{{\rm eff}}_{i-}\in(-\infty,X_{i0 })$ and $X^{{\rm eff}}_{i+}\in(X_{i0},\infty)$, where $X_{i0}=X_{i-}(b_0)=X_{i+}(b_0)$. For example, $X_{i0}=0$ in the example $i=m$
used in the previous Section, cf.~Eq.~(\ref{Xeffm}). Therefore, when performing the manipulations leading to (\ref{differentinnerprod}, we find for the 
cross term
\be
\int \dd\alpha_i\; e^{\im\frac{\alpha_i}{\plk} (X^{\rm eff}_{i+}-X^{\rm eff '}_{i-})}=0
\ee
except in the measure zero point $b=b_0$, killing the cross term.
The requirement that $X^{\rm eff}_i$ covers the real line is satisfied, but with the joint domains of $X^{{\rm eff}}_{i+}$ and $X^{{\rm eff}}_{i-}$ only, and without cross terms.  Therefore, for this inner product and in this approximation,
\bea
\langle\psi_1|\psi_2  \rangle&=&\prod_{i=1}^{D} \Big( \int \dd{X^{{\rm eff}}_{i+}}  \psi _{i+1}^\star (b,T_i) \psi_{i+2}(b,T_i) \nn\\
&&+\int \dd{X^{{\rm eff}}_{i-}}  \psi _{i-1}^\star (b,T_i) \psi_{i-2}(b,T_i)  \Big)
\label{approxnewinner}
\eea
and the interference between incident and reflected waves disappears. Moreover, the norm of a state only depends on the wave function in the classically allowed region. Calling this measure semiclassical therefore seems appropriate. 

In conclusion, for $b\ge b_0$ the probability in terms of $b$ has the 
form
\be
\label{probxeff}
{\cal P}_i(b;T_i)=|\psi_{i+}|^2\left| \frac{\dd X^{{\rm eff}}_{i+}}{\dd b}\right|+
|\psi_{i-}|^2\left| \frac{\dd X^{{\rm eff}}_{i-}}{\dd b}\right|\,.
\ee
For our model with radiation and $\Lambda$ and now assuming $k=0$ for simplicity, we have (cf.~Eq.~(\ref{xefflimits})) for the measure
factors
\be
\frac{\dd X^{{\rm eff}}_{1\pm}}{\dd b} = \frac{b^2}{2}\pm\frac{b^4-2m/\phi}{2\sqrt{b^4-b_0^4}}\,,\quad \frac{\dd X^{{\rm eff}}_{2\pm}}{\dd b} = \frac{\mp 1}{\sqrt{b^4-b_0^4}}\,.
\ee

In this semiclassical approximation, one can define an explicitly unitary notion of time evolution, focusing on one of the times $T_i$ and therefore on only one of the factors in (\ref{approxnewinner}). From the form of the inner product it is clear that a self-adjoint ``momentum'' operator is given by $-\im\plk\frac{\partial}{\partial X^{{\rm eff}}_{i\pm}}= -\im\plk\left( \frac{\dd X^{{\rm eff}}_{i\pm}}{\dd b}\right)^{-1}\frac{\partial}{\partial b}$, where in the first definition we think of $X^{{\rm eff}}_{i\pm}$ as a single variable going over the whole real line and in the second expression the sign depends on whether the operator acts on $\psi_{i+}$ or $\psi_{i-}$.

Moreover, the waves $\psi_{i+}$ are constructed to satisfy
\be
\im\plk\frac{\partial}{\partial T_i}\psi_{i\pm} = -\im\plk\frac{\partial}{\partial X^{{\rm eff}}_{i\pm}}\psi_{i\pm}\,,
\ee
see Eq.~(\ref{envelopes}) and the discussion below. Hence, they satisfy a time-evolution equation with a self-adjoint operator on the right-hand side, which is all that is needed.

\section{Towards phenomenology}\label{phenom}
One may rightly worry that our semiclassical inner product and other approximations have removed too much of the quantum behavior of the full theory.
For any state, the probability of being in the classically forbidden region would always be exactly zero. The phenomenon of ``ringing'' is erased. 
We need to go beyond the semiclassical measure and peaked wave-packet approximation to see these phenomena. And yet, even within these approximations we can infer some interesting phenomenology, which probably will survive the transition to a more realistic model~\cite{brunobounce}
 involving pressureless matter (rather than radiation) and $\Lambda$. We also refer to~\cite{brunobounce} for an investigation of effects 
revealed within the semiclassical approximation closer to the bounce than considered here. 

\begin{figure}
\includegraphics[scale=.9]{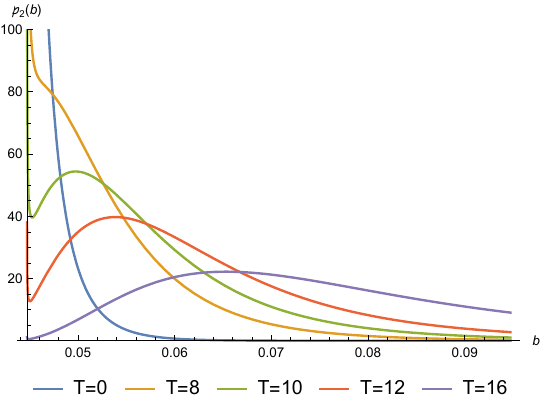}
\caption{Probability with the semiclassical measure, for the same situation as in Fig.\ref{reflectpsi}, at the various times $T_m=16,12,10,8,0$. We have explicitly verified that this probability density, unlike the function plotted in Fig.\ref{reflectprob}, always integrates to unity.}
\label{reflectprobmeasure}
\end{figure}

In Fig.~\ref{reflectprobmeasure}  we replot Fig.~\ref{reflectprob} using the semiclassical measure
(\ref{probxeff}) and Gaussian packets (\ref{Gauss}). Hence,
for the wave function factor associated with $m$ and $T_m$ we have
\bea
{\cal P}_2(b;T_m)&=&\frac{e^{-\frac{ (X_{+  2}^{\rm  eff} -T_m)^2}{2 \sigma_{T2} ^2}}
+e^{-\frac{ (X_{- 2}^{\rm  eff} -T_m)^2}{2 \sigma_{T2} ^2}}
}{\sqrt{2\pi\sigma^2_{T2}}\sqrt{b^4-b_0^4}}
\eea
without an interference term. At times well 
away from the bounce, the measure factor goes like $1/b^2$, so for a sufficiently
peaked wave packet it factors out. However, for times near the bounce the measure factor is significant. It induces a soft divergence
as $b\rightarrow b_0$,
 \be\label{Pb0}
{\cal P}_2(b\rightarrow b_0;T_m)=\frac{\exp{\left[-\frac{ T_m^2}{2 \sigma_{T2} ^2}\right]}}
{\sqrt{2\pi\sigma^2_{T2}}b_0^{3/2}\sqrt{b -b_0}}\,,
\ee
which becomes exponentially suppressed when $|T_m|\gg\sigma_{T2}$ (for example in Fig.~\ref{reflectprobmeasure} this is hardly visible
already for $T_m=16$), but is otherwise significant.
As we see in Fig.~\ref{reflectprobmeasure}, the measure factor therefore leads to a double-peaked distribution, when the main peak (due to the Gaussian) is present (in this picture at $T_m=10, 12, 16$). The measure factor also shifts the 
main peak of the distribution towards $b_0$, since it now follows
\be
T_m-X^{\rm eff}_{i\pm}=\mp \frac{2b^3\sigma_{Tm}^2}{b^4-b_0^4}\,,
\ee
which is valid for times when one of the waves dominates (incident or reflected), and the right-hand side is due fully to the measure effect.
We recall~\cite{JoaoPaper} that the classical trajectory is reproduced by $T_m=X^{\rm eff}_{i\pm}$. 
At some critical time close to the bounce, the ``main'' peak disappears altogether (see $T_m=8$ in  Fig.~\ref{reflectprobmeasure}),
with the distribution retaining a peak only at $b=b_0$. This peak becomes sharper and sharper as $|T_m|\rightarrow 0$
(so the average value of $b$ will eventually be larger than the classical trajectory, although the peak of the distribution will now be below 
the classically expected value, and stuck at $b_0$). A detailed study of how all these effects interact in a more concrete setting 
is discussed elsewhere~\cite{brunobounce}, but all of this points to interesting phenomenology near the $b$-bounce at $b_0$. 
The strength of the effects, and for how long they will be felt, depends on $\sigma_T$ for whichever clock is being used, which in turn depends
on the sharpness of its conjugate ``constant''. The sharper the progenitor constant, the larger the $\sigma_T$, and so the stronger the effect around the $b$-bounce.

How this fits in with other constraints pertaining to the life of the Universe well away from the $b$-bounce has to be taken into consideration. See, e.g., \cite{brunobounce} for a realistic model for which an examination of these details is more meaningful. We note that in real life it is the dominant clock for pressureless matter (rather than for radiation) that is relevant. This could be the same as the dominant clock for radiation (for example, if both are derived from a deconstantization of Newton's $G$; see~\cite{pad1,twotimes}) or not.

\section{Conclusions}\label{concs}
In this paper we laid down the foundation for studying the quantum effects of the bounce in $b$ which our Universe has recently experienced. 
We investigated a toy model designed to be simple whilst testing the main issues of a transition from deceleration to acceleration: 
a model with only radiation and $\Lambda$. The realistic case of a mixture of matter and $\Lambda$ is studied in~\cite{brunobounce}. Nonetheless, we were able to unveil both promising and disappointing results. 

Analogies with quantum reflection and ringing were found, but these will require going beyond the semiclassical approximation. 
Specifically, the inner product issues presented in Section~\ref{inner} were tantalizing in that they point to new quantum effects,
namely in the nonlocal nature of probability, as highlighted in Section~\ref{innermulti}. However, as soon as the semiclassical approximation is
consistently applied to both solutions and inner product, even the usual interference of incident and reflected waves is erased (see
Section~\ref{innerbounce}). 

Nonetheless, the semiclassical measure factor has a strong effect on the probabilities near the bounce, as was shown in Sections~\ref{innerbounce} and \ref{phenom}. It introduces a double-peaked distribution for part of the trajectory\fn{Strictly speaking the divergence at $b_0$ is always present, but at times well away from the bounce this is 
negligible; cf. Eq.~(\ref{Pb0}).}. This eventually becomes single peaked, with the average $b$ shifting significantly from the classical trajectory. 
The period over
which this could be potentially felt depends on the width of the clock, $\sigma_T$. This is not {\em a priori} fixed, since the concept of squeezing is not
well defined in a ``unimodular'' setting, as pointed out in~\cite{JoaoPaper}. Indeed, any deconstantized constant can be seen as the 
constant momentum of an abstract free particle moving with uniform ``speed'' in a ``dimension'' which we identify with a time variable.
It is well known that, unlike for a harmonic oscillator or electromagnetic radiation~\cite{knight}, coherent states for a free a particle lack a natural scale with which to define dimensionless quadratures and so the squeezing parameter~\cite{freecoh}. Hence, they share this problem with the free particle \fn{This is also found in the quantum treatment of the parity odd component of torsion appearing in first order theories~\cite{quantumtorsion}, or in any other quantum treatment of theories with trivial classical dynamics.}. Thus, an uncertainty in $T$ and $b$ of the order
of a few percent, felt over a significant redshift range around the bounce, is a distinct possibility. 
It is tempting to relate these findings to the so-called ``Hubble tension'' (see, e.g., Ref.~\cite{HubbleTension} and references therein), as is done in~\cite{brunobounce}.

It should be stressed that due to Heisenberg's uncertainty principle involving constants and conjugate times,
if we define sharper clocks (so that the fluctuations studied herein are not observable), it might be their conjugate constants that bear observable uncertainties.
This would invalidate the approximations used in this paper (namely, those leading to wave packets and the semiclassical measure). Most
crucially, 
$b_0$, the point of reflection, would not be sharply defined for such states, with different partial waves reflecting at different ``walls'' and then
interfering. 
Such quantum state for our current Universe should not be so easily dismissed. It might be an excellent example of 
cosmological quantum reflection.  
 
We close with two comments. In spite of its ``toy'' nature, our paper does make a point of principle: 
quantum cosmology could be here and now, rather than something swept under the carpet of the ``Planck epoch''. 
This is not entirely new (see, e.g., Ref.~\cite{QCnow}), but it would be good to see such speculations get out of the toy model
doldrums. Obviously, important questions of interpretation would then emerge~\cite{Jonathan,Jonathan1}. 
Finally, we note that something similar to the bounce  studied here happens in a reflection in the reverse direction at the end of inflation. 
One may wonder about the interconnection between any effects studied here and re-/preheating.

{\it Acknowledgments.} We would like to thank Bruno Alexandre, Jonathan Halliwell, Alisha Mariott-Best and Tony Padilla for discussions related to this paper. The work of SG was funded by the Royal Society through
a University Research Fellowship (UF160622) and a Research Grant (RGF$\backslash$R1$\backslash$180030). The work of JM was supported by the STFC Consolidated Grant ST/L00044X/1.

\end{document}